\documentclass[aps,twocolumn,epsfig,floatfix]{revtex4}

\usepackage[latin1]{inputenc}

\usepackage{amsmath,amsthm,amsmath,amsfonts,amssymb,color,graphicx,bbm}
\newtheorem{theorem}{Theorem}

\newtheorem{lemma}[theorem]{Lemma}

\def\Symp#1,#2,#3,#4.{\left[\left(\begin{array}{c}#1\\#2\end{array}\right),\left(\begin{array}{c}#3\\#4\end{array}\right)\right]}
\def\Vec#1,#2.{\left(\!\begin{array}{c}#1\\#2\end{array}\!\right)}
\def\tuple#1.{\langle#1\rangle}
\def\ket#1.{|#1\rangle}
\def\bra#1.{\langle#1|}
\def\braket#1,#2.{\langle#1|#2\rangle}
\def\Gauss#1,#2.{{#1\brack#2}_d}

\def\ZZ{\mathbbm{Z}}
\def\RR{\mathbbm{R}}
\def\CC{\mathbbm{C}}

\def\H{\mathcal{H}}

\DeclareMathOperator{\tr}{tr}

\DeclareMathOperator{\supp}{supp}
\DeclareMathOperator{\const}{const}

\begin{document}

\title{Non-negative Wigner functions in prime dimensions}

\author{D. Gross}

\affiliation{
	Institute for Mathematical Sciences, Imperial College London, London
	SW7 2BW, UK
} 
\affiliation{
	QOLS, Blackett Laboratory, Imperial College London, London SW7 2BW,
	UK
}

\email{david.gross@imperial.ac.uk}

\date{\today}

\begin{abstract}
According to a classical result due to Hudson, the Wigner function of
a pure, continuous variable quantum state is non-negative if and only
if the state is Gaussian. We have proven an analogous statement for
finite-dimensional quantum systems. In this context, the role of
Gaussian states is taken on by stabilizer states. The general results
have been published in [D.\ Gross, J.\ Math.\ Phys.\ {\bf 47}, 122107
(2006)]. For the case of systems of odd prime
dimension, a greatly simplified proof can be
employed which still exhibits the main ideas. The present paper gives
a self-contained account of these methods.
\end{abstract}

\pacs{}

\maketitle

\section{Introduction}

The Wigner distribution establishes a correspondence between quantum
mechanical states and real pseudo-probability distributions on phase
space. 'Pseudo' refers to the fact that, while the Wigner function
resembles many of the properties of probability distributions, it can
take on negative values. It is therefore of interest to characterize
those quantum states that are classical in the sense of giving rise
to non-negative phase space distributions.

For the case of pure states described by vectors in $\H=L^2(\RR)$, the
resolution of this problem was given by Hudson in Ref.\
\cite{hudson}, and later extended to multiple-particles by
Soto and Claverie (Ref.\ \cite{soto}).

\begin{theorem} (\emph{Hudson, Soto, Claverie)}
	Let $\psi\in L^2(\RR^n)$ be a state vector. The Wigner function of
	$\psi$ is non-negative if and only if $\psi$ is a \emph{Gaussian
	state}.
	
	By definition, a vector is Gaussian if and only if it is of the form
	\begin{equation*}
	%\label{posStates}
		\psi(q)=e^{2\pi i (q \theta q+x q)},
	\end{equation*}
	where $x\in\RR^n$ and $\theta$ is a symmetric matrix with entries in
	$\CC$
	\footnote{
		Note that the boundedness of $\psi\in L^2(\RR^n)$ implies that
		$\theta$ has positive semi-definite imaginary part. 
	}.
\end{theorem}

It is our objective to prove that the situation for discrete quantum
systems is very similar, at least when the dimension of the Hilbert
space is odd. The following Theorem states the main result. 

\begin{theorem} \emph{(Discrete Hudson's Theorem)}\label{thMain}
	Let $d$ be odd and $\psi\in L^2(\ZZ_d^n)$ be a state vector.  The
	Wigner function of $\psi$ is non-negative if and only if $\psi$ is a
	\emph{stabilizer state}.

	Given that $\psi(q)\neq0$ for all $q$, a vector $\psi$ is a
	stabilizer state if and only if it is of the form
	\begin{equation*}
		\psi(q)=e^{\frac{2\pi}{d} i (q \theta q+x q)},
	\end{equation*}
	where $x\in\ZZ_d^n$ and $\theta$ is a symmetric matrix with entries in
	$\ZZ_d$.
\end{theorem}

In the previous theorem, $\ZZ_d:=\{0,\dots,d-1\}$ denotes the set of
integers modulo $d$.

It turns out that, although the formulation of Hudson's result carries
naturally over to finite dimensional systems, the respective proofs
are radically different. The original argument relies crucially on
function theory, which is of course not available in the setting that
this paper addresses.

Recently, Galvao \emph{et.\ al.\ }took a first step into the direction
of classifying the quantum states with positive Wigner function (see
Ref.\ \cite{galvao}).  To explain the relationship of their results to
the present paper, we have to comment shortly on two different
approaches to defining discrete Wigner functions.
On the one hand, it has long been realized that Wigner's definition
carries over naturally to discrete odd-dimensional systems (Ref.\
\cite{longPaper,oldWootters,vourdas,leonhardt,miquel,villegar,klimov,ruzzi,
chaturvedi,diploma,wootters}
and Section \ref{scWigner}).
This approach is the one used in the present paper.
On the other hand, Gibbons, Hoffmann, and Wootters listed a set of
axioms that candidate definitions have to fulfill in order to resemble
the properties of the well-known continuous case (Ref.\
\cite{wootters}).  Let us call functions that fall into this class
\emph{generalized Wigner functions}. The characterization does not
specify a unique solution: for a $d$-dimensional Hilbert space, there
exist $d^{d-1}$ distinct generalized Wigner functions. The
construction of Gibbons \emph{et.\ al.\ } has been described only for the
case where $d$ is the power of a prime.

If the dimension of the Hilbert space is of the form $d^n$, a second
ambiguity arises.  We are free to conceive such a space either as
being associated to a system of $n$ constituents, each of dimension
$d$, or to a single one of dimension $d^n$. While the Wigner function
is the same for both cases, the set of stabilizer states is not (see
Refs. \cite{longPaper, diploma}).  Indeed, the 'single-particle'
stabilizer states turn out to be a proper subset of the
'multiple-particle' ones. As a striking example, the generalized Bell
and GHZ states nor are not stabilizer states on single $d^2$ or
$d^3$-dimensional systems. 

%\footenote{ 
	%The ratio of the cardinalities of the respective sets is given
	%roughly by $d^{\frac12n}$ 
%}

In Ref.\ \cite{galvao} it was proved that a state of a
\emph{single-particle} system of prime-power dimension is a stabilizer
state if and only if \emph{all} its generalized Wigner functions are
non-negative. The authors aim to establish necessary requirements for
quantum computational speedup. Indeed, if the Wigner function of a
quantum computer is positive at all times, then it operates only with
stabilizer states and hence offers no advantage over classical
computers, by the Gottesman-Knill Theorem (Ref.\ \cite{nielsen}).

For the case of pure states, our results imply the ones of Ref.\
\cite{galvao} and exceed them in two ways.  Firstly, we show that it
suffices to check positivity for a single definition of the Wigner
function, as opposed to $d^{d-1}$ ones. Secondly, our statements hold
for multiple-particle systems, which constitute the proper setting for
both quantum computation and the Gottesman-Knill Theorem. On the other
hand, Ref.\ \cite{galvao} makes assertions about mixed states and qubit
systems, which are not covered by our findings.

Our general results have been published in Ref.\ \cite{longPaper}.
However, the proof is rather involved. Many technicalities arise due
to the fact that for non-prime $d$, arithmetic modulo $d$ lacks the
desirable properties of finite fields. Our aim in writing Ref.\
\cite{longPaper} was to achieve the broadest possible generality in
spite of these difficulties. The downside of this approach is that
core ideas of the argument are obscured by technical issues. The
present paper employs a different method of proof, which is available
only for systems of odd prime dimension. For this special case, the
main result of Ref.\ \cite{longPaper} can be obtained using only a
fraction of the space. It is our hope that this paper makes the ideas
accessible to a wider audience.

The next section summarizes further findings contained in Ref.\
\cite{longPaper}.  We go on to recall the definition and properties of
discrete Wigner functions in Section \ref{scWigner}.  Section
\ref{scMainTheorem} is devoted to a complete proof of the easiest
special case of Theorem \ref{thMain}, that being given by a single
particle on a Hilbert space of prime dimension. 

\section{Further results and implications}

It is natural to ask how Hudson's results generalize to mixed states.
Certainly, mixtures of Gaussian states are positive on phase space and
Narcowich in Ref.\ \cite{narcowich} conjectured that all such quantum
states are convex combinations of Gaussian ones. Br\"ocker and Werner
refuted the conjecture  by giving a counter-example (Ref.\
\cite{werner}).  We show in Ref.\ \cite{longPaper} that the situation
is similar in the finite setting.

Further, we show how to lift the ambiguity in the axiomatic
characterization of Wigner functions by requiring Clifford covariance,
note that a unitary operator preserves positivity if and only it is a
Clifford operation, discuss the relation of various ways to introduce
Wigner functions and stabilizer states in dimensions of the form
$d=p^n$, and give an explicit account on the connection between
stabilizer states and Gaussian states.

\section{Wigner Functions}\label{scWigner}

This section provides a very superficial introduction to
discrete Wigner functions. We allow ourselves to refer the reader to
Refs. \cite{longPaper} for further details. In what follows $d$
denotes an odd prime. All integer arithmetic in this paper is
implicitly assumed to be $\bmod\, d$. The symbol $2^{-1}=(d+1)/2$ is
the multiplicative inverse modulo $d$. All state vectors are elements
of the Hilbert space $\H$ spanned by $\{\ket0.,\dots,\ket d-1.\}$.
Lastly, $\omega=e^{\frac{2\pi}d i}$ is a $d$th root of unity.

The relations
\begin{eqnarray*}\label{shiftClock}
  x(q)\ket k. = \ket k+q., \quad\quad
	z(p)\ket k. = \omega^{p k} \ket k.
\end{eqnarray*}
define the \emph{shift} and \emph{boost} operators respectively. The
most central element in the theory are the 
\emph{Weyl operators} (in quantum information also known as the
\emph{generalized Pauli operators}) given by
\begin{eqnarray*}
	w(p,q)=\omega^{-2^{-1}p q} z(p)x(q).
\end{eqnarray*}
The
\emph{characteristic function} of an operator $\rho$ is given by the expansion
coefficients of $\rho$ in terms of the Weyl operators
\begin{eqnarray*}
	\Xi_\rho(\xi,x) = \frac1d \tr(w(\xi,x)^\dagger \rho).
\end{eqnarray*}
We define the \emph{Wigner function} to be the symplectic
Fourier transform of the characteristic function:
\begin{eqnarray*}
	W_\rho(p,q) 
	&=& \frac1d \sum_{\xi,x \in \ZZ_d} \omega^{p\xi-q x}\, \Xi_\rho(\xi,x).
\end{eqnarray*}
It is a tedious yet straight-forward computation to show that the
Wigner function of a pure state is given by
\begin{eqnarray*}
	W_{\psi}(p,q) 
	&:=& W_{\ket\psi.\bra\psi.}(p,q) \\
	&=& 
	\frac1{d} \sum_{\xi \in \ZZ_d} \omega^{-\xi p}\,
	\psi(q+2^{-1} \xi)\bar\psi(q-2^{-1} \xi). \nonumber 
\end{eqnarray*}

If $S$ is a $2\times 2$-matrix with elements in $\ZZ_d$ and
determinant $1$, then there exists a unitary operation $\mu(S)$ (the
\emph{Weil} \cite{weil}
\footnote{
	There is a confusing similarity of names: the Weil representation
	(after Andr\'e Weil) acts on the Weyl operators (after Hermann Weyl).
} or \emph{metaplectic} representation of $S$) such that
\begin{equation*}
	\mu(S) w(p,q) \mu(S)^\dagger = w(S(p,q)).
\end{equation*}
The Wigner function is covariant in the sense that, if
$\rho'=\mu(S)\,\rho\,\mu(S)^\dagger$, then
\begin{equation}\label{weilCovariance}
	W_{\rho'}(p,q) = W_\rho(S(p,q)).
\end{equation}

Similarly, the Weyl operators induce translations of the Wigner
function. Letting $\rho'=w(p',q')\,\rho\,w(p',q')^\dagger$, it holds
that
\begin{equation}\label{weylCovariance}
	W_{\rho'}(p,q) = W_\rho(p+p',q+q').
\end{equation}

The \emph{Clifford group} is the set of unitary matrices that send
Weyl operators to Weyl operators under conjugation 
\footnote{
	Note that the ``Clifford group'' which appears in the context of
	quantum information theory \cite{gottesman} has no connection to the
	group by the same name used	e.g.\ in the representation theory of
	$SO(n)$.
}.
Every Clifford mapping is of the form $w(p,q)\mu(S)$
%for some $(p,q), S$.
and hence preserves positivity of the Wigner function.

Finally, \emph{stabilizer states} are the images of the computational
basis states under the action of the Clifford group.

\section{Main Theorem -- Single particles in prime dimensions}
\label{scMainTheorem}

Define the \emph{self correlation function} 
\begin{equation*}
	K_\psi(q,x)=\psi(q+2^{-1}x) \bar\psi(q-2^{-1}x)
\end{equation*}
and note that the Wigner function obeys
\begin{equation*}
	W(p,q)
	=\frac1{d} \sum_{x} \omega^{-p x} K_\psi(q,x) .
\end{equation*}
Recall that the Fourier transform $\hat f$ of a function $f: \ZZ_d \to
\CC$ is defined to be $\hat f(x)=1/d \sum_q \omega^{-qx} f(q)$.
Therefore, for a fixed $q_0$, $W(p,q_0)$ is the Fourier transform of
$K(q_0,x)$.  Hence $W$ is non-negative if and only if the $d$
functions $K(q_0,\cdot)$ have non-negative Fourier transforms.

In harmonic analysis, the set of functions with non-negative Fourier
transforms is characterized by a well-known theorem due to Bochner.  
%The following theorem states 
We state an elementary version of Bochner's
Theorem, along with a variation for subsequent use.

\begin{theorem}\label{thBochner}
	\emph{(Variations of Bochner's Theorem)}
	Consider a function $f: \ZZ_d\to \CC$. It holds that
	\begin{enumerate}
		\item 
		The Fourier transform of $f$ is non-negative if and only if the
		matrix
		\begin{equation*}
			{A^x}_q=f(x-q)
		\end{equation*}
		is positive semi-definite.

		\item
		The Fourier transform of $f$ has constant modulus 
		(i.e. $|\hat f(x)|=\const$)
		if and only if $f$ is orthogonal to its translations:
		%, that is, iff
		\begin{eqnarray*}
			\tuple f, \hat x(q) f. = \sum_{x} \bar f(x) f(x-q) = 0,
			%\quad q \neq 0.
		\end{eqnarray*}
		for all non-zero $q\in \ZZ_d$.
	\end{enumerate}
\end{theorem}

\begin{proof}
	The matrix $A$ is circulant. It is well-known that circulant
	matrices are normal (hence diagonalizable) with eigenvalues given by
	the Fourier transform of the first row (up to a positive
	normalization constant). The first claim is now immediate.

	By the same argument, $A$ is proportional to a unitary matrix if and
	only if $|\hat f(q)|$ is constant. But a matrix is unitary if and
	only if its rows form an ortho-normal set of vectors.
\end{proof}

The next three lemmas harvest some consequences of Bochner's Theorem
to gain information on the pointwise modulus $|\psi(q)|$ of the vector.

\begin{lemma} \label{lmIneq} 
	\emph{(Modulus Inequality)}
	Let $\psi$ be a state vector with positive Wigner function.

	It holds that
	\begin{equation*}
		|\psi(q)|^2 \geq |\psi(q-x)|\,|\psi(q+x)|
	\end{equation*}
	for all $q,x \in \ZZ_d$.
\end{lemma}

\begin{proof}
	Fix a $q \in \ZZ_d$.  As $W_\psi$ is non-negative, so is the Fourier
	transform of $K_\psi(q,x)$ with respect to $x$.  Bochner's Theorem
	implies that ${A^x}_y = K(x-y,q)$ is positive semi-definite
	(\emph{psd}) which in turn implies that all principal sub-matrices
	are psd. In particular the determinant of the $2\times 2$ principal
	sub-matrix 
	\begin{eqnarray*}
		&&
		\left(
		\begin{array}{cc}
			K_\psi(q,0) & K_\psi(q,2x) \\
			K_\psi(q, -2x) & K_\psi(q,0) 
		\end{array}
		\right) \\
		&=&
		\left(
		\begin{array}{cc}
			|\psi(q)|^2	&	\psi(q+x)\bar\psi(q-x) \\
			\bar\psi(q+x)\psi(q-x) & |\psi(q)|^2 
		\end{array}
		\right)
	\end{eqnarray*}
	must be non-negative. But this means
	\begin{eqnarray*}
		|\psi(q)|^4 - |\bar\psi(q+x)	\psi(q-x)|^2 \geq 0,
	\end{eqnarray*}
	which proves the theorem.
\end{proof}

We will call the set of points where a state-vector is non-zero its
\emph{support}.

\begin{lemma}\label{lmSupport}
	\emph{(Support Lemma)}
	Let $\psi$ be a state vector with positive Wigner function.

	If $\psi$ is supported on two points, then it has maximal support.
\end{lemma}

\begin{proof}
	Denote by $S=\supp\psi$ the support of $\psi$.  $S$ has the property
	to contain the midpoint of any two of its elements. Indeed, if
	$a, b \in S$, then setting $q=2^{-1} (a+b)$ and $x=2^{-1}(a-b)$ in
	the Modulus Inequality shows that
	\begin{equation*}
		|\psi(2^{-1}(a+b))| \geq |\psi(a)|\,|\psi(b)| > 0,
	\end{equation*}
	hence $2^{-1}(a+b) \in S$. 

	Assume there exist two points $a,b\in S$. 
	Requiring $a=0$ is no loss of generality, for else we substitute
	$\psi$ by $\psi'=w(0,-a)\psi$. By Eq.\ (\ref{weylCovariance}),
	$\psi'$ has positive Wigner function if and
	only if $\psi$ has.

	We claim that
	\begin{equation}\label{balancedProp}
		2^{-l}\beta \,b \in S
	\end{equation}
	for all $l$ and $\beta\leq2^l$. The proof is by induction on $l$.
	Suppose Eq.\  (\ref{balancedProp}) holds for some $l$. If
	$\beta \leq 2^{l+1}$ is even, then $2^{-l-1}\beta\,b=
	2^{-l}(\beta/2)b\in S$. Else, 
	\begin{eqnarray*}
		2^{-l-1} \beta\, b 
		= 
		2^{-1}\big( 2^{-l} \frac{\beta-1}2\, b + 2^{-l} \frac{\beta+1}2\,b \big)
		\in S,
	\end{eqnarray*} 
	which proves the claim.

	Now,by Fermat's Little Theorem $2^{d-1}=1 \mod d$ and hence, setting
	$l=d-1$ in Eq.\ (\ref{balancedProp}), we conclude that $\beta\, b \in
	S$ for all $\beta \leq d-1$. But every point in $\ZZ_d$ is of that
	form.
\end{proof}

\begin{lemma}\label{constlemma}
	\emph{(Constant Modulus)} 
	Let $\psi$ be a state vector with positive Wigner function and
	maximal support.

	%The modulus of $\psi$ is constant: $|\psi(x)|=\const$.
	Then $|\psi(q)|=\const$.
\end{lemma}

\begin{proof}
	Pick two points $x, q \in \ZZ_d$ and suppose $|\psi(q)|>|\psi(x)|$.
	
	Letting $z=x-q$, the assumption reads $|\psi(q)|>|\psi(q+z)|$.  Lemma
	\ref{lmIneq} centered at $q+z$ gives
	\begin{eqnarray*}
		|\psi(q+z)|^2 
		&\geq& |\psi(q)|\,|\psi(q+2z)| \\
		&>& |\psi(q+z)| \, |\psi(q+2z)|,
	\end{eqnarray*}
	therefore $|\psi(q+z)| > |\psi(q+2z)|$.  By inducting on this
	scheme, we arrive at
	\begin{equation*}
		|\psi(q)| > |\psi(q+z)| > |\psi(q+2 z)| > \cdots
		%|\psi(q)| > |\psi(q+1)| >  \cdots > |\psi(q+d)| = |\psi(q)|
	\end{equation*}
	and hence $|\psi(q)|>|\psi(q+dz)|=|\psi(q)|$, 
	which is a contradiction. 

	Thus $|\psi(q)|\leq|\psi(x)|$. Swapping the roles of $x$ and $q$
	proves that equality must hold.
\end{proof}

\begin{theorem}\label{mainPrime}
	\emph{(Main Theorem -- Special Case)}
	Let $d$ be prime and $\psi \in L^2(\ZZ_d)$ be a state vector with
	positive Wigner function.
 	Then $\psi$ is a stabilizer state.
\end{theorem}

\begin{proof}
	By the Support Lemma, $\psi$ is either a position eigenstate or
	else it has maximal support. In the former case, $\psi$ is
	manifestly a stabilizer state, so we need only treat the latter. 
	Let $U$ be a Clifford operation. Since $U$ preserves positivity,
	%maps positive Wigner functions to positive Wigner functions, 
	the Support Lemma applies to
	$U\psi$. Suppose $U$ is such that $\supp U\psi$ contains
	just a single point. Then $U\psi$ belongs to the computational
	basis and hence, by definition, $\psi$ is a stabilizer state.
	
	Therefore, we are left to treat those state vectors whose image
	under any Clifford operation has maximal support.  The proof is
	concluded by showing that such states do not exist.

	For assume there is such a vector $\psi$.  As $\psi$ has pointwise
	constant modulus, so does $K_\psi$. Employing Theorem \ref{thBochner},
	we find that, for every fixed $q_0$, $W(p,q_0)$ is orthogonal to its own
	translations. But since $W$ is non-negative, it follows that
	$W(p,q_0)$ can be non-zero on at most one point.
	A Wigner function that is concentrated at a single point can not
	represent a physical state
	\footnote{
		Such a Wigner function corresponds to a Hermitian operator with
		both positive and negative eigenvalues (see Ref.\
		\cite{longPaper}). One can think of this fact as an incarnation of
		the uncertainty principle.	
	}.
	There must hence exist at least two points $a, b$ in the support of
	$W$ (note that we are now considering the support of Wigner
	functions and no longer the support of state vectors). Making once
	more use of the fact that translations are implemented by Clifford
	operations, assume $a=0$.  There exists a unit-determinant matrix
	$S$ that sends $b$ to a vector of the form $Sb=(0,q_0)^T$. But then
	there are two points in the support of $W_{\mu(S)\psi}(p,q_0)$,
	contradicting our earlier derivation.
\end{proof}

\section{Summary}

We have proved a 'classicality result' for discrete Wigner functions:
those state vectors which give rise to a classical probability
distribution in phase space belong to the set of stabilizer states.
These, in turn, allow for an efficient classical description.
Comparing the proof of the special case treated here to the involved
argument employed in Ref.\ \cite{longPaper}, it becomes apparent how
much the geometrical properties of integer residues modulo prime
numbers simplify the structure.

\section{Acknowledgments}

The author is grateful for support and advice provided by Jens Eisert
during all stages of this project. Comments by and discussions with
K.\ Audenaert S.\ Chaturvedi, H.\ Kampermann, M.\ Kleinmann, A.\
Klimov, M.\ Ruzzi,  and C.K.\ Zachos are kindly acknowledged.

This work has benefited from funding provided by the European
Research Councils (EURYI grant of J. Eisert), the European Commission
(Integrated Project QAP), the EPSRC (Interdisciplinary Research
Collaboration IRC-QIP), and the DFG.

\end{document}